\documentclass[a4paper,fleqn]{cas-dc}

\usepackage[authoryear]{natbib}

\def\tsc#1{\csdef{#1}{\textsc{\lowercase{#1}}\xspace}}
\tsc{WGM}
\tsc{QE}

\begin{document}
\let\WriteBookmarks\relax
\def\floatpagepagefraction{1}
\def\textpagefraction{.001}

\shorttitle{}    

\shortauthors{Izzo et al.}  

\title{The Environment of High-Energy Transient Host Galaxies at High Redshifts as Observed by SHARP}

\author[1]{L. Izzo$^{1}$}

\affiliation{organization={INAF-Osservatorio Astronomico di Capodimonte, Napoli, Italy},
            addressline={salita Moiariello 16}, 
            city={Naples},
            postcode={80131}, 
            country={Italy}}

\cormark[1]

\fnmark[1]

\ead{luca.izzo@inaf.it}



\cortext[1]{Corresponding author}



\begin{abstract}
The study of transient events, such as Gamma-Ray Bursts (GRBs) and extreme Supernovae (SNe), provides a unique probe of star-forming environments in the early Universe. Traditional global host galaxy studies often fail to capture the local conditions around the progenitor, leading to potential mischaracterizations due to metallicity gradients and spatial inhomogeneities. Using high spatial resolution integral field unit (IFU) data, such as those obtained with SHARP/VESPER, allows us to spatially resolve the host galaxies and accurately determine the physical properties (metallicity, stellar age, kinematics) at the precise explosion site. Here, we review the need for near-infrared spatially-resolved observations to link progenitor properties, such as the required low metallicity for long GRBs, to their birth environments, and highlight the crucial role of instruments like SHARP in advancing high-redshift transient science, including the characterization of SNe without associated GRBs and Kilonovae hosts.
\end{abstract}


\begin{keywords}
 \sep \sep \sep
\end{keywords}

\maketitle

\section{Introduction}

\label{sec1}

Massive stellar explosions, including Long Gamma-Ray Bursts (LGRBs) and Super-Luminous Supernovae (SLSNe), are powerful tools for studying star formation and chemical evolution in high-redshift environments. The progenitors of these events, particularly LGRBs associated with broad-lined Type Ic Supernovae (Ic-BL SNe), are widely believed to require low-metallicity environments due to the need for chemically-stripped envelopes \citep{MacFadyen1999}. During the collapse of such a massive star, we witness the formation of a GRB when the jet produced by the central engine during the final core collapse, will break out of the rapidly rotating progenitor, the {\it Collapsars} scenario \citep{Woosley1993, MacFadyen1999}. 

GRB-SNe are typically found in the brightest regions of their host galaxies \citep{Kelly,Fruchter,Savaglio,Japelj}. This is expected because their massive, stripped-envelope progenitors have short lifetimes of only $\sim$10–30 Myr \citep{Maund}. Consequently, these stars explode close to the H II regions in which they formed, allowing for spatially resolved studies of both their immediate environments and neighbouring star-forming regions. Such analyses provide direct constraints on the physical conditions relevant to massive-star evolution, including stellar age, extinction, metallicity, and star-formation rate, as well as insights into whether these progenitors may have formed in binary systems.

The global host galaxy properties (e.g., integrated metallicity and stellar mass) often fail to represent the conditions directly surrounding the progenitor, especially in star-forming galaxies exhibiting steep metallicity gradients and chemical inhomogeneities \citep{Leloudas2015}.  This necessitates the use of spatially-resolved spectroscopy to obtain accurate, localized measurements. The metallicity-mass diagram clearly shows that GRB hosts tend toward the low-metallicity, low-mass regime, but spatially-resolved studies are needed to confirm the local environment.

\section{Spatially-Resolved Host Galaxy Analysis}

Integral Field Spectroscopy (IFS) offers the capability to map the distribution of stellar and gas properties across the host galaxy. This approach allows for a direct comparison of the physical properties at the transient location with the global galaxy characteristics.

\subsection{GRB Host Galaxy Studies}

Recent work on LGRB and SLSN hosts \citep{Kruhler,Izzo1,Thone,Izzo2}, demonstrates the power of spatially-resolved data. By applying Stellar Population Synthesis (SPS) models \citep{BC03,STARLIGHT,MILES,FSPS,PPXF} to both the global host spectrum and the local HII regions, we can recover:
\begin{itemize}
    \item The star-formation history (SFH) and stellar metallicity (e.g., fitting stellar continuum models).
    \item Gas-phase physical properties, using the SPS best-fit to recover clean emission line fluxes for diagnostics.
\end{itemize}

These localized analyses are critical. For instance, in the case of GRB 100316D \citep{Izzo1}, emission-line diagnostics place the explosion site firmly within a star-forming HII region, allowing for a precise determination of the gas-phase metallicity and ionization mechanism (e.g., BPT diagram analysis \citep{BPT}). The stellar population analysis further shows a complex SFH, suggesting that localized HII region properties are not accurately represented by the galaxy's integrated light.

Moving at high redshifts, deep space-based imaging has revealed the presence of multiple systems surrounding the location of the majority of GRB events. A clear example is provided by the distant GRB 210905A at z = 6.3 \citep{Saccardi}, where the use of spatially-resolved spectroscopy would definitely allow us to study the multiple systems located in the vicinity of this event, see Fig. 1.

\subsection{The Puzzle of Ic-BL SNe without GRBs}
Only a small fraction ($<8\%$) of all observed Ic-BL SNe are associated with a detected LGRB. This raises fundamental questions about the nature of the majority of Ic-BL events. Potential explanations include: 1) Off-axis GRBs: The relativistic jet is successfully launched but is pointed away from the Earth, meaning the $\gamma$-rays are missed, but the broad-lined supernova is observed; 2) Choked Jets: The jet fails to fully penetrate the progenitor's stellar envelope, resulting in a successful supernova explosion but no observed $\gamma$-ray burst. 

Investigating the host environments of Ic-BL SNe without GRBs is crucial to constrain the metallicity threshold required for jet production. Current IFU surveys of nearby Ic-BL hosts (e.g., SN 2009bb, SN 2014ad, see Fig. 2, Izzo et al. in prep.) show a range of metallicities (e.g., $12+\log(\text{O/H}) \sim 8.1$ to $8.5$), reinforcing the need for highly accurate, spatially-resolved data to determine if a hard metallicity boundary exists and whether it is a primary driver in GRB production. Moreover, the currently available sample of Ic-BL SN host galaxies is largely limited to optical observations at low redshift. Future observations with SHARP could extend similar analyses to higher redshifts, significantly increasing the number of accessible host galaxies and enabling more systematic studies.

\subsection{Probing the Cold Gas Phase at High-z}
GRB afterglow spectra provide an additional, powerful probe of the interstellar medium (ISM) in the host galaxy, specifically the neutral, cold gas phase. GRB afterglows penetrate the multiple absorbing clouds in the star-forming regions. Studies like GRB 220627A at $z \approx 3.084$ (with intervening systems at $z \approx 2.665$) allow the measurement of absorption line metallicities, providing constraints on the cold gas metallicity ($[\text{X/H}]$) in the host and foreground systems \citep{DeWet2023}. Crucially, in cases where both absorption (neutral/cold) and emission (ionized/hot) metallicities are available, there is often good agreement (within $\pm0.4$ dex, \citep{Schady}), suggesting a similar composition of these phases at high redshift. This agreement is vital for establishing robust metallicity diagnostics at high redshift based solely on emission lines, which are often easier to obtain.

\section{The Role of SHARP and High-z Metallicity}
The {\bf S}pectroscopy for {\bf H}igh-z {\bf A}rcs and {\bf R}andom {\bf P}opulations (SHARP) instrument, particularly with the VESPER upgrade, is optimally positioned to execute targeted, spatially-resolved surveys of transient hosts.

The SHARP wavelength coverage ($\Delta \lambda_{SHARP}=[0.95-2.45]\mu m$) provides access to key rest-frame emission lines over a wide range of redshifts:
\begin{itemize}
    \item $\text{H}\alpha$ and $\text{[O}\,\text{III]}$ at $z < 2.7$.
    \item $\text{[O}\,\text{II]}$ at $z < 5.5$, enabling $\text{SFR}$ and metallicity determination.
\end{itemize}
The high signal-to-noise ratio achievable for these emission lines, even for faint hosts, makes SHARP an ideal facility for:
\begin{itemize}
    \item Building a targeted sample of spatially-resolved LGRB hosts out to $z \approx 3$, providing accurate metallicities at the GRB location.
    \item Investigating the metallicity evolution for relativistic SNe (with or without GRBs), crucial for testing the metallicity-dependent occurrence rate of Core-Collapse SNe (CCSNe) \citep{Pessi2023}.
\end{itemize}
The recent JWST detection of an Ic-BL at $z=2.83$ (SN 2023adt, \citep{Siebert2024} underscores the importance of obtaining high-quality spectroscopy for high-redshift transients, a regime where SHARP can provide necessary complementary data.

Based on MUSE data published in De Wet et al. (2023), we extrapolated the spectrum to near-infrared wavelengths assuming a young (0.1 Gyr) stellar population modeled with Bruzual \& Charlot templates at sub-solar metallicity. Using this model, we simulated the exposure time required to achieve a signal-to-noise ratio (SNR) of 10 at the rest-frame [O II] 3727, 3729 Å lines, which are commonly used for metallicity and star formation rate studies. Assuming the VESPER mode and a spatial resampling of 7×7, we find that the [O II] emission can be detected with a total exposure time of less than 1.5 hours. This SNR, combined with the spectral resolution of VESPER (R $\approx$ 3000), should also enable the identification of narrow absorption features from the host galaxy’s interstellar medium, allowing estimates of column densities and, consequently, the metallicity of the cold gas component. Exposure time calculations were performed using the SHARP Exposure Time Calculator (ETC v0.6).

The spectrum of GRB 220627A reported in De Wet et al. (2023) was extracted from a comparable region in the ESO/MUSE dataset, although it covers a different rest-frame wavelength range than that accessible with SHARP. For a similar spatial extraction, the average SNR achieved with MUSE is approximately a factor of two lower than that expected with SHARP, assuming a spatial resampling of 5×5. This comparison highlights the potential of SHARP for studies of transient host galaxies at moderately high redshift, and suggests a complementary role as a near-infrared extension to future MUSE and BLUEMUSE observations. 

If we consider space-based facilities such as the James Webb Space Telescope (JWST), the G140M disperser provides a spectral resolution comparable to that of SHARP, although nominally somewhat lower. Without spatial rebinning and assuming a similar input spectrum, we obtain a slightly higher SNR, which can increase by up to a factor of four with modest spatial binning for a one-hour exposure. JWST delivers superior sensitivity compared to SHARP, but having access to SHARP would strengthen the feasibility of such programs.

\begin{figure*}
  \centering
  \includegraphics[width=16cm]{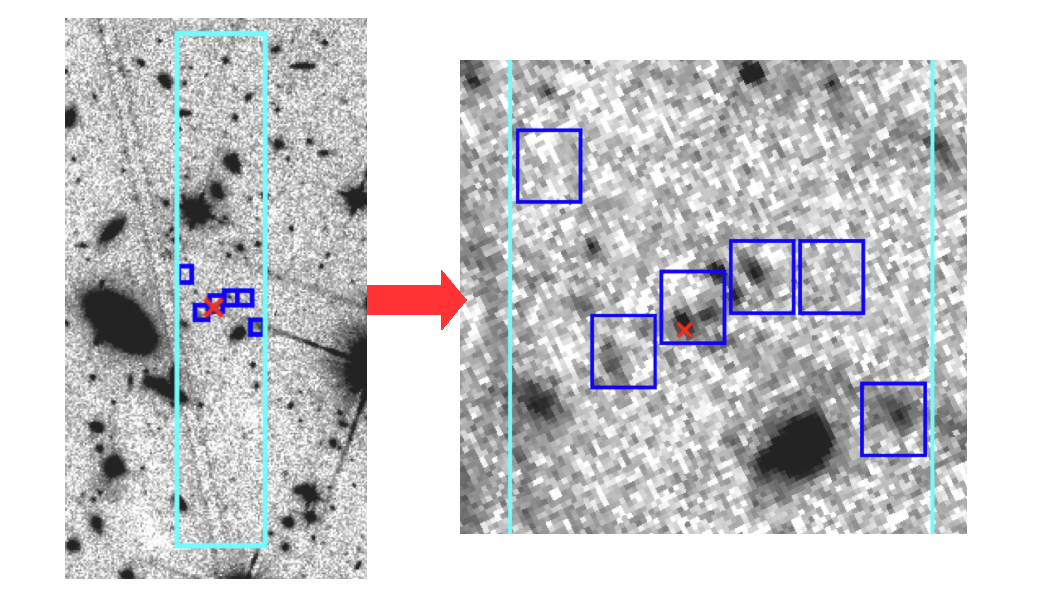}
    \caption{An example of SHARP/VESPER usage is shown for the host system of GRB 210905A \citep{Saccardi}. The ability to position VESPER’s mini-IFUs on individual bright galaxies surrounding the GRB event enables targeted studies of their physical properties through emission- and absorption-line analysis.}\label{fig0}
\end{figure*}

\begin{figure*}
  \centering
  \includegraphics[width=16cm]{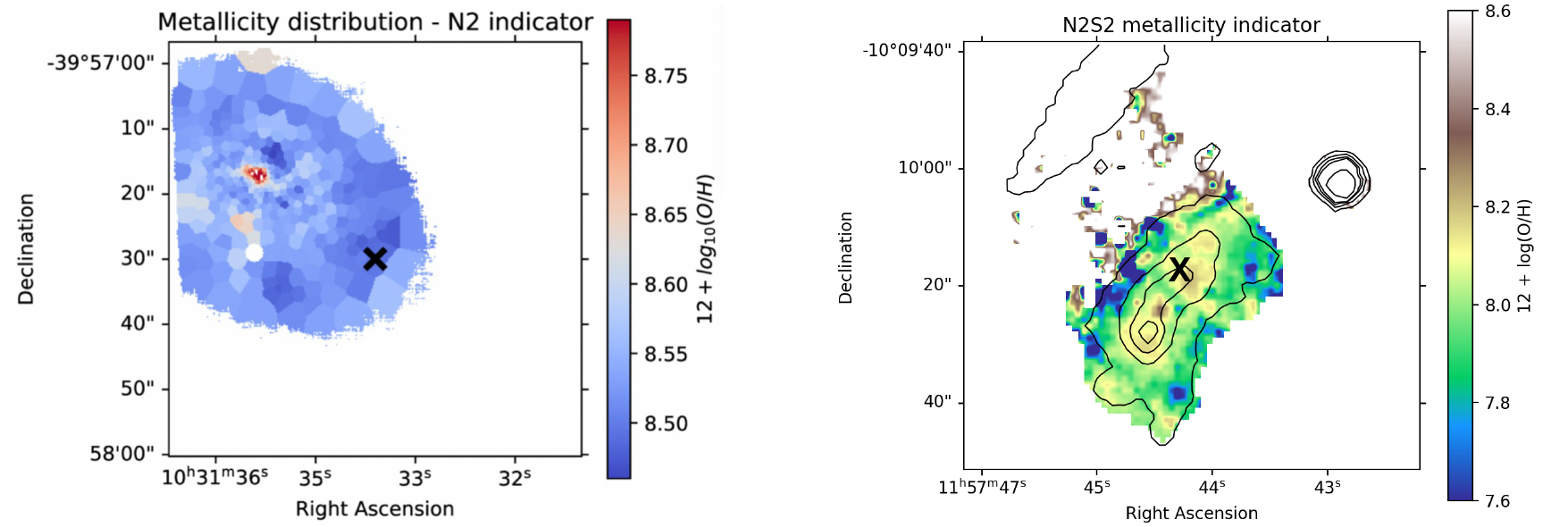}
    \caption{The spatially-resolved metallicity, obtained with emission line indicators, for the host galaxies of two Ic-BL SNe, namely SN 2009bb and SN 2014ad (Izzo et al. in prep.). The exact position in the host of both SNe is indicated with a black cross.  Note that this figure is partially based on results from work in preparation and is included here for illustrative purposes only to show the potential usage of SHARP as a tool to study the immediate environment of astrophysical transients at relatively high-redshifts.}\label{fig1}
\end{figure*}


\section{Kilonovae and Future Synergy}
The study of Kilonovae (KNe) and short GRBs (e.g., AT 2017gfo/GW170817 and GRB 230307A) also benefits from detailed host environment characterization. These events are often associated with binary mergers and older stellar populations.

\subsection{Kilonova Host Analysis}
In the case of GW170817/AT 2017gfo, the host galaxy NGC 4993 showed an old stellar population at the KN position ($\sim 5-12$ Gyr), suggesting a massive binary system progenitor with a complex evolutionary history \citep{Hjorth,Levan1}. More recently, the host of GRB 230307A, observed with MUSE and JWST/NIRCAM, was found to be a relatively low-mass galaxy ($M_{*} \sim 10^9 M_{\odot}$) dominated by an old, sub-solar metallicity population, with the KN located $\sim 40$ kpc offset from the young star-forming component \citep{Levan2}, see also Fig. 3. This confirms that progenitors of short GRBs/KNe often have large offsets from active star-forming regions and arise from older stellar populations. SHARP can provide IFU coverage for a statistical sample of these offset events, probing the low-surface-brightness outskirts of host galaxies where KNe occur.

\subsection{Synergy with Large Surveys}
Future transient searches in the deep fields of surveys like the Vera C. Rubin Observatory's LSST \citep{Izevic} and the Euclid mission \citep{Euclid} will significantly increase the sample size of detected transients up to $z \approx 2.5$. The combined high-quality optical (LSST) and NIR (Euclid, Roman) data will allow for robust photometric classification and host property determination. SHARP will provide the essential spectroscopic follow-up, offering unique, spatially-resolved metallicity and kinematic maps for the most peculiar and high-redshift events discovered by these surveys. The planned transient search in Euclid Deep Fields, in synergy with LSST, promises to build a large statistical sample of SNe with excellent multi-wavelength coverage.

\begin{figure*}
  \centering
  \includegraphics[width=16cm]{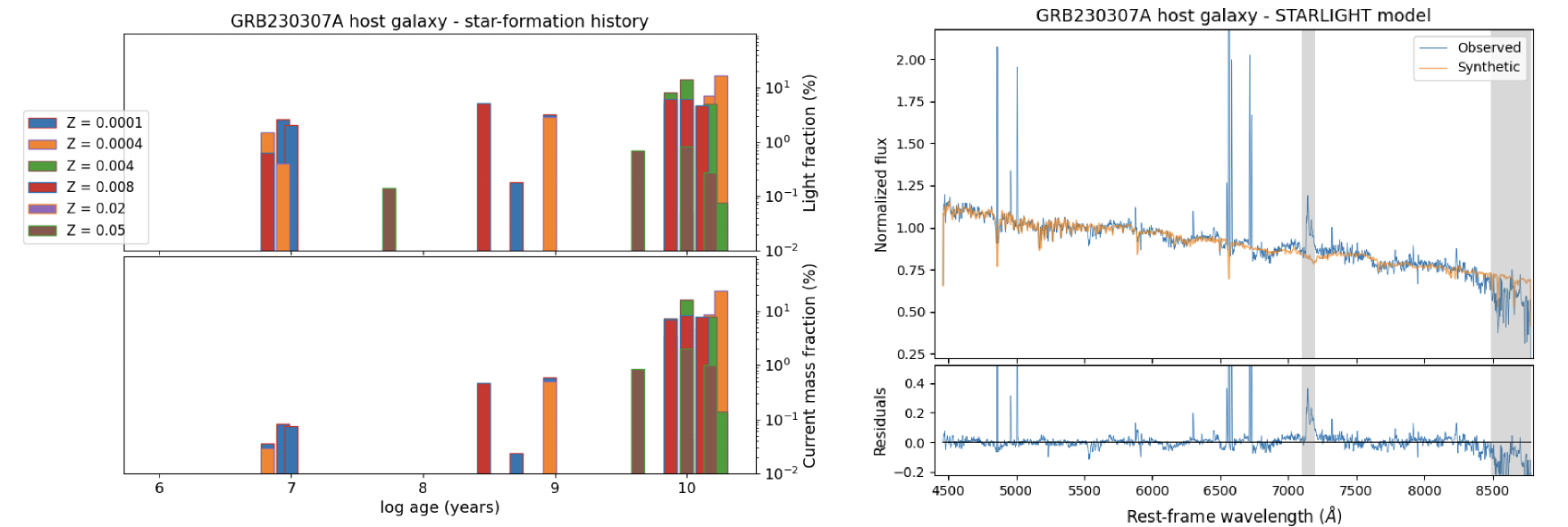}
    \caption{The fit performed using STARLIGHT \citep{STARLIGHT} of the integrated light, reconstructed from the ESO-MUSE datacube, of the host galaxy of GRB 230307A. The resulting star-formation history derived using a Bruzual \& Charlot stellar template library, and a Chabrier initial mass function, suggests that the galaxy is dominated by a main old stellar population with sub-solar metallicity and  a younger component, likely originating from inflow/accretion of pristine gas.}\label{fig2}
\end{figure*}

\section{Conclusion}
The era of transient astronomy requires moving beyond galaxy-integrated measurements to spatially resolved studies. The SHARP/VESPER instrument is a crucial tool for this effort, allowing astronomers to directly link the properties of transient progenitors (e.g., the low metallicity required for LGRBs, the older stellar populations for KNe) to the physical conditions of their immediate birthplaces at high redshift. Future work, in synergy with large photometric surveys like LSST and Euclid, promises to provide a complete picture of the environments where the most energetic stellar explosions in the Universe occur.






\printcredits




\end{document}